\documentclass[10pt,a4paper,final]{iopart}
\expandafter\let\csname equation*\endcsname\relax
\expandafter\let\csname endequation*\endcsname\relax
\usepackage[margin=1 in]{geometry}
\usepackage{amsmath,color,iopams,graphicx}
\usepackage[breaklinks=true,colorlinks=true,linkcolor=blue,urlcolor=blue,citecolor=blue]{hyperref}
\usepackage[english]{babel}
\usepackage{graphicx,comment}
\usepackage{mathtools}
\usepackage{amsmath}
\usepackage{amsfonts}
\usepackage{amssymb}
\usepackage{mathrsfs}
\usepackage{footnote}
\makesavenoteenv{tabular}
\makesavenoteenv{table}
\usepackage{multicols}
\usepackage{subfigure}
\usepackage{widetext}
\usepackage{caption}
\begin{document}
\title{Tunneling density of states of fractional quantum Hall edges: an unconventional bosonization approach}
\author{  Nikhil Danny Babu$^{\dagger}$, Girish S. Setlur$^{*}$}
\address{Department of Physics \\ Indian Institute of Technology  Guwahati \\ Guwahati, Assam 781039, India}
\ead{$^{\dagger}$danny@iitg.ac.in,$^{*}$gsetlur@iitg.ernet.in}

%------------------------------------ABSTRACT now-------------------------------------------------------------------------------------------------------
\begin{abstract}
An unconventional bosonization approach that employs a modified Fermi-Bose correspondence is used to obtain the tunneling density of states (TDOS) of fractional quantum Hall (FQHE) edges in the vicinity of a point contact.  The chiral Luttinger liquid model is generally used to describe FQHE edge excitations. We introduce a bosonization procedure to study edge state transport in Laughlin states at filling $\nu = 1/m$ with $m$ odd (single edge mode) in the presence of a point contact constriction that brings the top and bottom edges of the sample into close proximity. The unconventional bosonization involves modifying the Fermi-Bose correspondence to incorporate backscattering at the point contact, leaving the action of the theory purely quadratic even in presence of the inhomogeneity. We have shown convincingly in earlier works that this procedure correctly reproduces the most singular parts of the Green functions of the system even when mutual forward scattering between fermions are included. The most singular part of the density-density correlation function (DDCF) relevant to TDOS calculation is computed using a generating functional approach. The TDOS for both the electron tunneling as well as the Laughlin quasiparticle tunneling cases is obtained and is found to agree with previous results in the literature. For electron tunneling the well-known universal power laws for TDOS viz. $ \sim \mbox{  }\omega^{ m-1 }$ and for quasi-particle tunneling the power law $ \sim \mbox{  } \omega^{ \frac{1}{m}-1 } $ are both correctly recovered using our unconventional bosonization scheme. This demonstrates convincingly the utility of the present method which unlike conventional approaches, does not treat the point-contact as an afterthought and yet remains solvable so long as only the most singular parts of the correlation functions are desired.
\end{abstract}

\vspace{2pc}
\section{Introduction}
In one dimension, electron interactions have a drastic effect and a Fermi liquid description of such a system fails \cite{10.1063/1.1704046,PhysRevB.9.2911,PhysRevLett.33.589}. Instead it was found that the 1D interacting system exhibits a Luttinger liquid phase \cite{FDMHaldane_1981}. The Luttinger liquid exhibits only collective low energy excitations moving to the right and left. It was shown by Wen \cite{PhysRevB.43.11025} that the fractional quantum Hall effect (FQHE) \cite{PhysRevLett.48.1559} edge excitations could be described in terms of 1D interacting electrons. In fact it was shown that the edge states of a FQHE bulk with filling factor  $\nu = 1/m$ with $m$ an odd integer (termed the Laughlin series \cite{PhysRevB.23.5632}) can be described as chiral Luttinger liquids. The Laughlin FQHE states have only a single edge excitation and the top edge or bottom edge can be considered equivalent to the right moving or left moving half of a conventional Luttinger liquid. The FQHE edge states are the most experimentally accessible systems where one encounters Luttinger liquid physics. The FQHE edge states are immune to impurity backscattering and the low energy properties cannot be inferred by probing the bulk. But inter edge tunneling can be made possible by bringing the opposite edges of the sample together by forming a point contact. This is achieved by forming a constriction in the bulk through the application of a gate voltage. The low energy physics of the edge states can be probed by studying the tunneling transport properties through the point contact. The point contact in a $\nu = 1/m$ Hall fluid is isomorphic to a point impurity in a conventional one-channel Luttinger liquid. The transport properties through the point contact depends on the tunneling density of states (TDOS) of each edge. Power law suppression of TDOS is characteristic of a Luttinger liquid.\\
The proper way to deal with interactions in one dimensional systems is using bosonization. This involves replacing fermions with bosonic degrees of freedom. But conventional or standard bosonization is ill suited to be applied to systems with impurity backscattering. Therefore the common method to study transport through the point contact is by a perturbative treatment of the impurity \cite{doi:https://doi.org/10.1002/9783527617258.ch4}. As an alternative one can treat the impurity exactly by modifying the standard Fermi-Bose correspondence itself to take into account the impurity backscattering. This radical new method involves treating the Fermi-Bose correspondence as a mnemonic to obtain the correct correlation functions rather than a strict operator identity. This allows one to compute the most singular parts of the interacting Green's functions in presence of an impurity in one dimension and this technique has proven to be very successful in studying Luttinger liquids with impurities \cite{Danny_Babu_2020,doi:10.1142/S0217751X18501749,das2019nonchiral,DAS2017216,Das_2018,DAS20193149,Das_2020}. In the present article we extend this idea to FQHE edge states with a point contact constriction at the origin. We show that the universal power law in the TDOS is recovered using the modifed bosonization procedure in the presence of a point contact in both the cases of electron tunneling as well as quasiparticle tunneling. The paper is organized as follows. In Sec.\ref{model} we introduce the Hamiltonian of the model we will be discussing. In Sec.\ref{nonintgreen} we discuss the nonequilibrium Green's functions for the model in the absence of interaction i.e. for edge states of an integer quantum hall effect (IQHE) sample with $\nu=1$ and with the presence of a point contact driven out of equilibrium. In Sec.\ref{nonintboson} the modified Fermi-Bose correspondence for noninteracting edge states in presence of interedge tunneling through a point contact is discussed. In Sec.\ref{fqheboson} we set up the formalism of the proposed modified unconventional bosonization method for Laughlin FQHE edge states in the presence of a point contact. In Sec.\ref{ddcf} we compute using the generating functional method the density-density correlation function (DDCF) relevant to the TDOS calculation. In Sec.\ref{tdos} the tunneling density of states for electron tunneling and quasiparticle tunneling is calculated and the correct power law exponents are obtained. In Sec.\ref{linearcurrent} we calculate the current in response to a potential difference between the edges without a point contact using the DDCF obtained in Sec.\ref{ddcf} and obtain the expected fractional conductance result. We summarize our results and discuss the future extensions of our method in Sec.\ref{conclusion}.
\section{Model Hamiltonian}
\label{model}
We consider a system of two chiral Luttinger liquids with a point-contact tunnel junction at the origin. The non-interacting part of the Hamiltonian is
\begin{align}
H_{0} = \sum_{p}(v_F p + e V_{b})c^{\dagger}_{p,R}c_{p,R} &+ \sum_{p}(-v_Fp)c^{\dagger}_{p,L}c_{p,L}+ \frac{\Gamma}{L}(c^{\dagger}_{.,R}c_{.,L}+c^{\dagger}_{.,L}c_{.,R})
 \label{eqhamnonint}
\end{align}
where $R$ and $L$ label the right and left moving chiral spinless modes and $V_{b}$ is the bias voltage applied to the right movers. The fermion creation and annihilation operators in momentum space are $c^{\dagger}_{p}$ and $c_{p}$ respectively and we use the notation $c^{\dagger}_{.,R} = \sum_{p}c^{\dagger}_{p,R}$. We consider a symmetric point-contact junction with tunneling amplitude $\Gamma$ and the $L$ that does not appear in the subscript is the system size. We define the bias potential as $\mu_{L}-\mu_{R} = -\mu_{R} = -e V_{b} = e V$ following the same convention as in \cite{PhysRevB.93.085440}. The coupling at the point contact is not treated as a weak perturbation in our approach but is treated exactly as an impurity of arbitrary strength. The right mover and left mover fermions are coupled through a density-density interaction governed by $v_{0}$,
\begin{align}
H_{int} = \int dx\mbox{ } v_{0}\mbox{ } \rho_{R}(x,t) \rho_{L}(x,t)
\label{hint}
\end{align}
This system of chiral Luttinger liquids is a good model for the edge states of fractional quantum Hall systems, at least the ones with a single edge mode as in the Laughlin series with filling fraction $\nu = 1/m$ with $m$ an odd integer. We will show how this connection is made in the subsequent sections. The full Hamiltonian of interest is
\begin{align}
H\mbox{ }=\mbox{ }H_{0} + H_{int}
\end{align}
\section{Nonequilibrium Green functions for $H_{0}$}
\label{nonintgreen}
The noninteracting system was solved exactly in \cite{Babu_2022} and the full nonequilibrium Green functions were obtained as,
\begin{align}
 <\psi^{\dagger}_{\nu^{'}}(x^{'},t^{'})\psi_{\nu}(x,t)>_{0} \mbox{ }=\mbox{ } -\frac{i}{2\pi} \frac{\frac{\pi}{\beta v_{F}}}{\sinh( \frac{ \pi }{\beta v_F } (\nu x-\nu^{'}x^{'}-v_F(t-t^{'}) ) )} \kappa_{\nu,\nu^{'}}
\label{eqnoneq}
 \end{align}
 where $\nu$,$\nu^{'} = \pm 1$ with $R=1$ and $L=-1$ and
 \begin{widetext}
 \small
 \begin{align}
 \kappa_{1,1} \mbox{ }=\mbox{ } \bigg(  U(t^{'},t) \left[1
-      \theta(x^{'})    \mbox{          }   \frac{  2 \Gamma^2
}{\Gamma ^2 +4 v_F^2}\right]&\mbox{          }    \left[1
-      \theta(x )    \mbox{          }   \frac{  2 \Gamma^2
}{\Gamma ^2 +4 v_F^2}\right]
  \nonumber \\ &+ \left( \frac{\Gamma }{v_F} \mbox{          } \frac{(2 v_F)^2
}{\Gamma ^2 +4 v_F^2}\right)^2 \mbox{          }
  \theta(x )     \theta(x^{'} )   \mbox{          } U(t^{'},t^{'} - \frac{x^{'}}{v_F})   \mbox{     }
   U(t - \frac{x}{v_F},t)   \bigg)
   \label{eqkappa1}
 \end{align}
 \begin{align}
 \kappa_{-1,-1} \mbox{ }=\mbox{ } \bigg( \left[ 1
-      \theta(-x^{'})    \mbox{          }   \frac{  2 \Gamma^2
}{\Gamma ^2 +4 v_F^2}
\right] &  \left[ 1
-      \theta(-x )    \mbox{          }   \frac{  2 \Gamma^2
}{\Gamma ^2 +4 v_F^2}
\right]   \nonumber \\ &+   \left( \frac{\Gamma }{v_F} \mbox{          } \frac{(2 v_F)^2
}{\Gamma ^2 +4 v_F^2}\right)^2 \mbox{          }  \theta(-x )    \theta( -x^{'} )    \mbox{          }
 U(t^{'} + \frac{x^{'}}{v_F},t + \frac{x}{v_F} ) \bigg)
 \label{eqkappa2}
 \end{align}
 \begin{align}
 \kappa_{1,-1} \mbox{ }= \mbox{ }\bigg( -U(t - \frac{x}{v_F},t)   \mbox{          }  &\left[ 1
-      \theta(-x^{'})    \mbox{          }   \frac{  2 \Gamma^2
}{\Gamma ^2 +4 v_F^2}
\right]
    \theta(x )   \mbox{          } \nonumber \\ &+
 U(t^{'} + \frac{x^{'}}{v_F},t)
\mbox{          }
\left[1
-      \theta(x )    \mbox{          }   \frac{  2 \Gamma^2
}{\Gamma ^2 +4 v_F^2}\right] \theta( -x^{'} )   \bigg) i  \frac{\Gamma }{v_F}\frac{(2 v_F)^2
}{\Gamma ^2 +4 v_F^2}
\label{eqkappa3}
 \end{align}
 \begin{align}
 \kappa_{-1,1} \mbox{ }=\mbox{ }\bigg( - U(t^{'},t + \frac{x}{v_F} ) &\left[1
-      \theta(x^{'})    \mbox{          }   \frac{  2 \Gamma^2
}{\Gamma ^2 +4 v_F^2}\right]  \theta( -x )
 \nonumber \\ &+     U(t^{'},t^{'} - \frac{x^{'}}{v_F})   \mbox{          }     \left[ 1
-      \theta(-x )    \mbox{          }   \frac{  2 \Gamma^2
}{\Gamma ^2 +4 v_F^2}
\right]  \theta(x^{'} )     \bigg) \mbox{          } i \frac{\Gamma }{v_F}    \frac{ (  2  v_F )^2   }{\Gamma ^2 +4 v_F^2}
\label{eqkappa4}
 \end{align}
 \end{widetext}
 where $U(\tau,t) \equiv e^{ - i \int _{ (t-t^{'})}^t  e V_{b}(s)ds }$ with an arbitrary time-dependent voltage bias and $\theta(x)$ is the Dirichlet regularized step function (this means $ \theta(x > 0) = 1, \theta(x < 0) = 0, \theta(0) = \frac{1}{2} $). The symbol $<\mbox{ }>_{0}$ means the correlations in the absence of fermion-fermion interactions. For the rest of this article we will assume a constant bias. \\
 The density-density correlation functions for a constant bias are computed using Wick's theorem,
  \begin{widetext}
 \begin{align}
  <&T \mbox{     } \rho_R(x,t) \rho_R(x^{'},t^{'}) >_0 -<\rho_R(x,t) >_0<\rho_R(x^{'},t^{'}) >_0\mbox{          }
 \nonumber \\ &=  \mbox{          }<T \mbox{     } \rho_L(-x,t) \rho_L(-x^{'},t^{'}) >_0-<\rho_L(-x,t) >_0<\rho_L(-x^{'},t^{'}) >_0 \mbox{          }\nonumber \\
&=\mbox{ } \left[    \frac{i}{2\pi} \frac{ \frac{ \pi }{\beta v_F } }{\sinh( \frac{ \pi }{\beta v_F } (x-x^{'}-v_F(t-t^{'}) ) )  } \right]^2 \nonumber \\ &
\mbox{          }\mbox{          }\mbox{          }\mbox{          }\mbox{          }
    \bigg(   \left[1
-      \theta(x^{'})    \mbox{          }   \frac{  2 \Gamma^2
}{\Gamma ^2 +4 v_F^2}\right]^2\mbox{          }    \left[1
-      \theta(x )    \mbox{          }   \frac{  2 \Gamma^2
}{\Gamma ^2 +4 v_F^2}\right]^2
  + \left( \frac{\Gamma }{v_F} \mbox{          } \frac{(2 v_F)^2
}{\Gamma ^2 +4 v_F^2}\right)^4 \mbox{          }
  \theta(x )     \theta(x^{'} )  \nonumber \\
  &\mbox{ }\mbox{          }\mbox{          }\mbox{          }+
\mbox{          }
\left[1
-       \frac{  2 \Gamma^2
}{\Gamma ^2 +4 v_F^2}\right]^2 \mbox{          }   \left( \frac{\Gamma }{v_F} \mbox{          } \frac{(2 v_F)^2
}{\Gamma ^2 +4 v_F^2}\right)^2 \mbox{          } \theta(x )     \theta(x^{'} )   \mbox{          }
   \left(   e^{ -i e V_b (t- t^{'} - \frac{x}{v_F} + \frac{x^{'}}{v_F}) }  +   e^{ -i e V_b (t^{'}-t  - \frac{x^{'}}{v_F}+ \frac{x}{v_F}) }
  \right)\bigg)
  \label{rhorhoRRLL}
 \end{align}
 and
 \begin{align}
 <&T \mbox{     } \rho_R(x,t) \rho_L(-x^{'},t^{'}) >_0-<\rho_R(x,t)>_0< \rho_L(-x^{'},t^{'}) >_0 \mbox{          }
 \nonumber \\ &=  \mbox{          }<T \mbox{     } \rho_L(-x,t) \rho_R(x^{'},t^{'}) >_0 - <\rho_L(-x,t) >_0<\rho_R(x^{'},t^{'}) >_0\mbox{          }\nonumber \\
 &= \mbox{ }\left[  \frac{i}{2\pi} \frac{ \frac{ \pi }{\beta v_F } }{\sinh( \frac{ \pi }{\beta v_F } (x-x^{'}-v_F(t-t^{'}) ) )  }\right]^2 \nonumber \\ &
 \mbox{          }\mbox{          }\mbox{          }\mbox{          }\mbox{          }
 \bigg(\left( i  \frac{\Gamma }{v_F}\frac{(2 v_F)^2
}{\Gamma ^2 +4 v_F^2} \right)^2
\mbox{          }
  \left( -    \left[ 1
-      \theta(x )    \mbox{          }   \frac{  2 \Gamma^2
}{\Gamma ^2 +4 v_F^2}
\right]^2    \theta(x^{'})  -
 \left[1
-      \theta(x^{'} )    \mbox{          }   \frac{  2 \Gamma^2
}{\Gamma ^2 +4 v_F^2}\right]^2    \theta( x)      \right) \nonumber \\
&\mbox{          }\mbox{          }\mbox{          }\mbox{          }+  \left( i  \frac{\Gamma }{v_F}\frac{(2 v_F)^2
}{\Gamma ^2 +4 v_F^2} \right)^2
\mbox{          }
  \left(e^{ -i e V_b (t^{'}-t - \frac{x^{'}}{v_F} + \frac{x}{v_F}) }    +e^{ -i e V_b (t-t^{'} - \frac{x}{v_F} + \frac{x^{'}}{v_F}) }   \right)   \mbox{          }     \left[ 1
-       \frac{  2 \Gamma^2
}{\Gamma ^2 +4 v_F^2}
\right]^2  \theta(x)
  \theta( x^{'} )\bigg)
  \label{rhorhoRLLR}
 \end{align}
 \end{widetext}
The density-density correlations satisfy the following identities,
\begin{align}
 <&T \mbox{     } \rho_{\nu}(\nu \mbox{ }x,t) (\rho_{\nu}(\nu\mbox{ }x^{'},t^{'})+\rho_{-\nu}(-\nu \mbox{ }x^{'},t^{'})) >_0 -<\rho_{\nu}(x,t) >_0<(\rho_{\nu}(\nu \mbox{ }x^{'},t^{'})+\rho_{-\nu}(-\nu\mbox{ }x^{'},t^{'})) >_0\mbox{          }=
 \mbox{          }\nonumber \\
  <&T \mbox{     }(\rho_{\nu}(\nu \mbox{ }x,t) + \rho_{-\nu}(-\nu \mbox{ }x,t)  ) \rho_{\nu}(\nu \mbox{ }x^{'},t^{'}) >_0 - <(\rho_{\nu}(\nu \mbox{ }x,t) + \rho_{-\nu}(-\nu \mbox{ }x,t) ) >_0<\rho_{\nu}(\nu\mbox{ }x^{'},t^{'}) >_0 \nonumber \\ &\mbox{     }=  \mbox{          }
  \left[    \frac{i}{2\pi} \frac{ \frac{ \pi }{\beta v_F } }{\sinh( \frac{ \pi }{\beta v_F } (x-x^{'}-v_F(t-t^{'}) ) )  } \right]^2
  \label{denidentity}
 \end{align}
 where $\nu$ take values $\pm 1$ with $1$ denoting $R$ (right movers) and $-1$ denoting $L$ (left movers). This noninteracting model is isomorphic to edge states of an integer quantum Hall effect (IQHE) \cite{PhysRevB.23.5632,doi:https://doi.org/10.1002/9783527617258.ch4} system coupled through a point contact. In \cite{Babu_2022} the tunneling current and conductance across the point contact is calculated from the Green functions. The tunneling current is defined usually as the rate of change of the difference in the number of right and left movers,
 \begin{align}
 I_{tun}(t) = e\mbox{ }\partial_{t}\frac{\Delta N}{2} = e\frac{i}{2}\left[H,\Delta N\right] = e\frac{i}{2}\left[H,N_{R}-N_{L}\right]
 \label{tundef}
 \end{align}
 This is evaluated as
 \begin{align}
  I_{tun}(t) \mbox{        } = \mbox{          }-i e \Gamma \mbox{        } \lim_{t^{'} \rightarrow t } \bigg( < \psi^{\dagger}_R(0,t^{'})  \psi_L(0,t) > - < \psi^{\dagger}_L(0,t)  \psi_R(0,t^{'}) > \bigg)
  \label{eqitun}
 \end{align}
 This is calculated using the nonequilibrium Green functions and we get
 \begin{align}
 I_{tun}(t) \mbox{        } = \mbox{          }
     \mbox{          }    \frac{   4 t_{p}^{2}  }{(t_{p}^{2} +1)^{2}}
\mbox{    }
\frac{e^2}{h } V
\label{eqtunc}
 \end{align}
 where $t_{p}$ is a tunneling parameter defined as $\Gamma = 2 v_{F} t_{p}$ and the differential tunneling conductance is
 \begin{align}
 G = \frac{d I_{tun}}{d V(\tau)} \mbox{ }=\mbox{ } G_{tun} \mbox{        } = \mbox{          }  \frac{   4 t_{p}^{2}  }{(t_{p}^{2} +1)^{2}}
\mbox{    }
\frac{e^2}{ h }
\label{eqcond}
 \end{align}
in accordance with the predictions of standard scattering theory \cite{LANDAUER198191,PhysRevB.31.6207}.
 \section{Bosonization of the noninteracting edge states with interedge tunneling}
 \label{nonintboson}
 In this section we show the IQHE edge can be bosonized in the presence of backscattering due to a point contact that brings the opposite edges into close proximity. Although the bosonized description is not necessary for the IQHE edges, it is of use since it can be generalized to the case of fractional quantum Hall effect (FQHE) edges where fermion interactions need to be taken into account. So in this section our Hamiltonian is $H = H_{0}$. Let us consider an IQHE edge with only one edge mode i.e. one filled Landau level. One can linearize the low energy states near the Fermi momentum and it is possible to express the low energy degrees of freedom in terms of bosons. The Fermi surface consists of two Fermi points at $+k_{F}$ and $-k_{F}$. We are dealing with chiral fermions here i.e. right movers ($k_{F}$) and left movers ($-k_{F}$). Let us consider the edge with right mover fermions (i.e. with momentum $k_{F}$). We may define the edge density (fluctuation) operator
 \begin{align}
 \rho(x) = : \psi^{\dagger}(x) \psi(x) :
 \end{align}
  It is normal ordered with respect to the filled Fermi sea. The Fourier transform of the density is
  \begin{align}
  \rho_{R}(x) = \frac{1}{L}\sum_{q} \rho^{R}_{q} e^{-i q x}
\end{align}
where $\rho^{R}_{q} = \sum_{k} :c^{\dagger}_{k,R} c_{k+q,R} :$ . The commutator of the densities can be shown to be
 \begin{align}
 [\rho^{R}_{q},\rho^{R}_{q^{'}}]
 &= \frac{q L}{2 \pi}\delta_{q+q^{'},0}
 \end{align}
 Simlarly for the edge with left movers it is $[\rho^{L}_{q},\rho^{L}_{q^{'}}] = -\frac{q L}{2 \pi}\delta_{q+q^{'},0}$. The commutator is zero for densities of opposite chiralities. Hence the commutation relation between the density operators is bosonic like. Thus we can construct bosonic annihilation and creation operators of the form
 \begin{align}
 b_{q} = \sqrt{\frac{2 \pi}{ L |q|}}\sum_{\chi}\theta(\chi q)\rho_{\chi}(q)
 \end{align}
and
\begin{align}
b^{\dagger}_{q} = \sqrt{\frac{2 \pi}{ L |q|}}\sum_{\chi}\theta(\chi q)\rho_{\chi}(-q)
\end{align}
where $\chi$ can take values $R\mbox{ }(+1)$ and $L\mbox{ }(-1)$. These operators satisfy the usual bosonic commutation relations $[b_{q},b_{q^{'}}] = 0 $, $[b^{\dagger}_{q},b^{\dagger}_{q^{'}}] = 0$ and $[b_{q}, b^{\dagger}_{q^{'}}] =  \delta_{q,q^{'}}$. So we may write the right mover density operator in terms of these bosonic operators as
\begin{align}
\rho_{R}(q) = \sqrt{\frac{  L q}{2 \pi}}b_{q}
\end{align}
or
\begin{align}
\rho_{R}(-q) = \sqrt{\frac{ L q}{2 \pi}}b^{\dagger}_{q}
\end{align}
The commutator of the densities in real space is
\begin{align}
[\rho_{R}(x), \rho_{R}(x^{'})] = \sum_q  \frac{q }{2\pi L }  e^{ -i q (x-x^{'}) }  =  \frac{i}{2\pi}   \delta^{'}(x-x^{'})
\end{align}
It is convenient to introduce a new field $\phi$ defined by
\begin{align}
\phi_{\chi}(x,t) = 2 \pi \int^{x}dy \mbox{    } \mbox{     } \rho_{\chi}(y,t)
\end{align}
where $\chi$ can be $R$ or $L$. It can be shown that the $\phi$ operators obey a Kac Moody commutation relation
\begin{align}
[\phi_{R}(x,t), \phi_{R}(x^{'},t)] &= -[\phi_{L}(x,t), \phi_{L}(x^{'},t)] \nonumber\\&= -i \pi \mbox{ }  sgn(x-x^{'})
\end{align}
In conventional bosonization the chiral Fermi fields are expressed as \cite{doi:https://doi.org/10.1002/9783527617258.ch4}
\begin{align}
 \psi_{\chi}(x,t) = e^{ 2 \pi i \mbox{ }\chi\int^{x} \rho_{\chi}(y,t) dy }
 \label{conventionalbose}
\end{align}
where $\chi = \pm 1$ for $R$ or $L$ respectively. The Fermi-Bose correspondence in Eq.\ref{conventionalbose} is proved by resorting to the basis states of homogeneous (translation-invariant) systems. Edge states are generally insensitive to disorder. Backscattering becomes possible only when the edges of opposite chirality are brought close together as in a point contact and inter-edge tunneling can take place. When we have an impurity in the system like the point-contact tunnel junction present in the model under consideration, the number of right movers and left movers are not separately conserved, hence bosonization using the conventional approach is not suitable. \\
We propose a method to handle this issue by modifying the standard Fermi-Bose correspondence in Eq.\ref{conventionalbose} to include the effects of backscattering of fermions from the point contact impurity. The modified Fermi-Bose correspondence takes the form,
 \begin{align}
 \psi_{\chi}(x,t) = e^{ 2 \pi i \chi  \int^{x} (\rho_\chi(y,t)+\lambda \mbox{ }\rho_{-\chi}(-y,t)) dy }
 \label{ncbt}
 \end{align}
 where the value of $\lambda$ which is either $0$ or $1$ dictates the absence or presence of the additional $\rho_{-\nu}(-y,t)$ term respectively. This is similar to the non-chiral bosonization technique (NCBT) that has been used to obtain the most singular parts of the Green's functions of interacting Luttinger liquids with impurities \cite{Danny_Babu_2020,doi:10.1142/S0217751X18501749}. It has been shown that the series expansion of the NCBT Green's functions in powers of fermion-fermion interaction strength matches term by term with standard fermionic perturbation theory (most singular terms). It has also been shown that the NCBT Green's functions with forward scattering between fermions satisfy the (most singular parts of the) exact Schwinger-Dyson equations \cite{das2019nonchiral}. These results are a clear indication that this formalism is not mere phenomenology.
 In \cite{babu2023unconventional} it is shown how one can construct a bosonization ansatz using Eq.\ref{ncbt} and reproduce the nonequilibrium Green functions for this noninteracting problem. The two point correlations are recovered using
 \begin{align}
 <T \mbox{ }\psi^{\dagger}_{\chi}(x,t) \psi_{\chi}(x^{'},t^{'})> \mbox{ }\sim\mbox{ } < e^{ -2 \pi i \chi  \int^{x} (\rho_\chi(y,t)+ \rho_{-\chi}(-y,t)) dy }\mbox{ }e^{ 2 \pi i \chi  \int^{x^{'}} \rho_\chi(y^{'},t^{'}) dy^{'} }>
 \end{align}
 or equivalently
 \begin{align}
 <T \mbox{ }\psi^{\dagger}_{\chi}(x,t) \psi_{\chi}(x^{'},t^{'})> \mbox{ }\sim\mbox{ } <e^{ -2 \pi i \chi  \int^{x} \rho_\chi(y,t) dy }\mbox{ } e^{ 2 \pi i \chi  \int^{x^{'}} (\rho_\chi(y^{'},t^{'})+ \rho_{-\chi}(-y^{'},t^{'})) dy^{'} }>
 \end{align}
 The choice of taking $\lambda=1$ for both the Fermi operators while evaluating the two-point correlations has been shown to be invalid as it doesn't obey the point-splitting constraint (please refer to Eq.23 of \cite{das2019nonchiral}).
 In a homogeneous system all but the lowest nontrivial moment (second moment) of the density vanish identically. In a system with impurity all odd moments of the density vanish identically but none of the even moments do. We make the crucial assertion that dropping all but the second moment in an inhomogeneous system amounts to studying the most singular parts of the Green functions (the proof of which is shown in \cite{Danny_Babu_2020}). Hence the Green functions are evaluated using the truncated version of the cumulant expansion
\begin{align}
<e^{A} e^{B}> \sim e^{\frac{1}{2}<A^{2}>} e^{\frac{1}{2}<B^{2}>} e^{<AB>}
\end{align}
 The piecewise constant prefactors are fixed by comparing with the exact solution obtained in \cite{Babu_2022}.
\section{Bosonizing the FQHE edge states with interedge tunneling}
\label{fqheboson}
A FQHE fluid with filling factor $\nu = 1/m$ with $m$ an odd integer (Laughlin series) has only a single edge mode. The right (left) edge state is equivalent to the right (left) half of a conventional Luttinger liquid. A point contact constriction in the fluid is isomorphic to an impurity barrier in a 1D Luttinger liquid. It follows from Wen's \cite{PhysRevB.43.11025,PhysRevLett.64.2206} hydrodynamic approach to describe the low-energy physics of the fractional quantum Hall edge states that the edge density operators have the following commutation relation
\begin{align}
[\rho^{\chi}_{q},\rho^{\chi}_{q^{'}}] = \chi \nu \frac{L q}{2 \pi}\delta_{q+q^{'},0}
\end{align}
where $\nu$ is the filling factor. This is pretty much same as the corresponding relation for the ordinary Luttinger liquid except for the additional factor of $\nu$. Hence it follows that the Kac Moody algebra of the $\phi$ fields picks up an extra $\nu$ factor.
\begin{align}
[\phi_{R}(x,t), \phi_{R}(x^{'},t)] &= -[\phi_{L}(x,t), \phi_{L}(x^{'},t)] \nonumber\\&= -i \mbox{ }\pi \mbox{ } \nu \mbox{ } sgn(x-x^{'})
\end{align}
So the operator for an edge excitation is written down with the following Fermi-Bose correspondence
\begin{align}
 \psi_{\chi}(x,t) = e^{i \chi \Phi/\nu} = e^{ 2 \pi i \mbox{ }\chi \frac{1}{\nu}\int^{x} \rho_{\chi}(y,t) dy }
\end{align}
This operator generally has fractional statistics and is not exactly fermionic but for the special case of $\nu = 1/m$ with $m$ odd this excitation is fermionic with charge $e$ and it represents an edge electron. In order to consider the effect of backscattering from a point contact between the edges we modify the Fermi-Bose correspondence as
\begin{align}
\psi_{\chi}(x,t) =  e^{ 2 \pi i \chi \frac{1}{\nu} \int^{x} (\rho_\chi(y,t)+\lambda \mbox{ }\rho_{-\chi}(-y,t)) dy }
\label{ncbtmod}
\end{align}
The idea is that Eq.\ref{ncbtmod} should be seen as a mnemonic that gives the most singular parts of the Green's functions for FQHE edge states with a point contact at the origin.
We need to determine the density-density correlation functions in presence of interactions before we can evaluate the two-point functions. In the next section we show how to evaluate these correlations using the generating functional method.
\section{Generating functional for density density correlations with interactions}
\label{ddcf}
Let us start by defining the following symmetric and antisymmetric density operators,
\begin{align}
 &\rho_{sym}(x,t) \equiv \rho_{R}(x,t)+\rho_{L}(-x,t) \mbox{ };\nonumber \\
 &\rho_{asy}(x,t) \equiv \rho_{R}(x,t)-\rho_{L}(-x,t)
 \label{rsymnonint}
 \end{align}
 This means we can write,
 \begin{align}
 \rho_{R}(x,t) = \frac{\rho_{sym}(x,t)+\rho_{asy}(x,t) }{2}
 \end{align}
 and
 \begin{align}
 \rho_{L}(-x,t) = \frac{\rho_{sym}(x,t)-\rho_{asy}(x,t) }{2}
 \end{align}
 Using Eqs.\ref{rhorhoRRLL},\ref{rhorhoRLLR} and \ref{denidentity} we write down the noninteracting correlations of the symmetric and antisymmetric densities,
 \begin{align}
 <\rho_{sym}(x,t)\rho_{sym}(x^{'},t^{'})>_{0} \mbox{ }=\mbox{ }2\left[    \frac{i}{2\pi} \frac{ \frac{ \pi }{\beta v_F } }{\sinh( \frac{ \pi }{\beta v_F } (x-x^{'}-v_F(t-t^{'}) ) )  } \right]^2
 \label{rsymrsym}
 \end{align}
\begin{align}
 <\rho_{sym}(x,t)\rho_{asy}(x^{'},t^{'})>_{0} \mbox{ }=\mbox{ } 0
 \end{align}
 \begin{align}
 <\rho_{asy}(x,t)\rho_{sym}(x^{'},t^{'})>_{0} \mbox{ }=\mbox{ } 0
 \end{align}
 \begin{align}
 <\rho_{asy}(x,t)\rho_{asy}(x^{'},t^{'})>_{0} \mbox{ }=\mbox{ }&\frac{-((4 v_{F}^{2} + \Gamma^{2})^{2}-32 v_{F}^{2} \Gamma^{2} \theta(x))((4 v_{F}^{2} + \Gamma^{2})^{2}-32 v_{F}^{2} \Gamma^{2} \theta(x^{'}))}{2 v_{F}^{2} \beta^{2}(4 v_{F}^{2} + \Gamma^{2})^{4}\sinh( \frac{ \pi }{\beta v_F } (x-x^{'}-v_F(t-t^{'}) ) )^{2}}\nonumber \\ &-\frac{64(-4 v_{F}^{3} \Gamma + v_{F} \Gamma^{3})^{2} \cos(\frac{- e V_{b}(x-x^{'}-v_F(t-t^{'}) )}{v_{F}})\mbox{ }\theta(x)\theta(x^{'})}{2 v_{F}^{2} \beta^{2}(4 v_{F}^{2} + \Gamma^{2})^{4}\sinh( \frac{ \pi }{\beta v_F } (x-x^{'}-v_F(t-t^{'}) ) )^{2} }
 \label{rhoasy}
 \end{align}
In the absence of the bias ($V_{b} = 0$) the noninteracting correlations satisfy the following relation
\begin{align}
  <T\rho_{asy}(x,t)\rho_{asy}(x^{'},t^{'})>_{0,\mbox{ }V_b = 0}\mbox{ }
  = \mbox{ }(\theta(xx^{'}) + r_1 \mbox{  }\theta(-xx^{'})) \mbox{   }<T\rho_{sym}(x,t)\rho_{sym}(x^{'},t^{'})>_{0}
  \label{nobiasrelation}
  \end{align}
   where we define $r_1 = \left(1
  -\frac{32 \Gamma ^2 v_F^2}{\left(\Gamma ^2+4 v_F^2\right)^2}
\right)$.
We choose to work with the $\rho_{sym}(x,t)$ and $\rho_{asy}(x,t)$ fields as it proves to be convenient in calculating the path integrals of the generating functional. The interaction Hamiltonian can now be expressed as
\begin{align}
H_{int} = \int dx \mbox{       } v_0 \mbox{     } \rho_R(x,t) \rho_L(x,t)  =\frac{ v_0 }{4}\mbox{     }  \int dx \mbox{       }  (\rho_{sym}(x,t)+\rho_{asy}(x,t) )
 ( \rho_{sym}(-x,t)-\rho_{asy}(-x,t) )
\end{align}
We introduce the auxiliary fields $U_{sym}$ and $U_{asy}$ and write down the generating functional
 \begin{align}
 Z[U] = \int D[\rho_{sym} ] \int D[\rho_{asy}]\mbox{ } \mbox{    }  e^{ i S_0 }e^{ i S_{int} } e^{ \int \rho_{sym} U_{sym} +  \int \rho_{asy} U_{asy} }
 \label{zu}
 \end{align}
 where $S_{0}$ and $S_{int}$ are the actions in absence and presence of interactions respectively, hence $e^{ i S_{int} }  =  e^{ -i \int_C dt \mbox{          }  H_{int} }$. The generating functional in the absence of interactions is
 \begin{align}
 Z_0[U] = &\int D[\rho_{sym} ] \int D[\rho_{asy}]\mbox{ } \mbox{    }  e^{ i S_0 } e^{ \int \rho_{sym} U_{sym} +  \int \rho_{asy} U_{asy} }\nonumber \\
 &\Rightarrow e^{ i S_0 } = \int D[U^{'}_{sym} ] \int D[U^{'}_{asy}]\mbox{ } \mbox{    }  e^{ -\int \rho_{sym} U^{'}_{sym} -  \int \rho_{asy} U^{'}_{asy} }\mbox{    } \mbox{    } Z_0[U^{'}]
 \label{z0}
 \end{align}
 Hence we may write
  \begin{align}
 Z[U] =  &\int D[U^{'}_{sym} ] \int D[U^{'}_{asy}]\mbox{    } Z_0[U^{'}] \mbox{ } \mbox{    } \int D[\rho_{sym} ] \int D[\rho_{asy}]\nonumber \\&e^{ -i \int_C dt \mbox{          }   \int dx  \mbox{ } \frac{ v_0 }{4}\mbox{     }   (\rho_{sym}(x,t)+\rho_{asy}(x,t) )
 ( \rho_{sym}(-x,t)-\rho_{asy}(-x,t) )   }
\mbox{      }\nonumber \\&e^{  \int_C dt \mbox{          }   \int dx  \mbox{ }  \rho_{sym}(x,t)( U_{sym}(x,t) - U^{'}_{sym}(x,t) )  +  \int_C dt \mbox{          }   \int dx  \mbox{ } \rho_{asy}(x,t) ( U_{asy}(x,t) - U^{'}_{asy}(x,t)) }
\label{zu}
 \end{align}
 The path integrals are evaluated using the saddle point method which involves calculating the extremum of the log of the integrand. After integrating over $\rho_{sym}$ and $\rho_{asy}$ we get
 \begin{align}
 Z[U] = &\int D[U^{'}_{sym} ] \int D[U^{'}_{asy}]\mbox{    } Z_0[U^{'}]
 \label{zuonly}
\end{align}
\[
 \times\mbox{ }e^{\substack{ -\frac{1}{i v_{0}} \int_C dt \mbox{  }   \int dx\mbox{ }( U_{asy}(-x,t) - U^{'}_{asy}(-x,t) - U^{'}_{sym}(-x,t) + U_{sym}(-x,t))(U_{asy}(x,t) - U^{'}_{asy}(x,t) + U^{'}_{sym}(x,t) - U_{sym}(x,t))}}
   \]
We make the most singular truncation of $Z_{0}$ in the RPA (random phase approximation) sense by using the Gaussian approximation
 \begin{align}
 Z_0[U^{'}] \mbox{ }  = \mbox{    }&e^{ \frac{1}{2} \int_C dt \int dx \int_C dt^{'} \int dx^{'} \mbox{     }<T\rho_{sym}(x,t)\rho_{sym}(x^{'},t^{'})  >_0 \mbox{      }   U^{'}_{sym}(x,t)U^{'}_{sym}(x^{'},t^{'})  }
\mbox{    }\nonumber \\&e^{ \frac{1}{2} \int_C dt \int dx \int_C dt^{'} \int dx^{'} \mbox{     }<T\rho_{asy}(x,t)\rho_{asy}(x^{'},t^{'})  >_0 \mbox{      }   U^{'}_{asy}(x,t)U^{'}_{asy}(x^{'},t^{'})  }
\label{z0u}
 \end{align}
 We have neglected all higher moments of $\rho$ in $Z_{0}[U^{'}]$ as they are less singular than the second moment \cite{Danny_Babu_2020} and therefore only the quadratic moment is included. Solving for the saddle point of the $U^{'}$ fields we get
 \begin{align}
 \int dt^{'} \int dx^{'} \mbox{     }<T\rho_{sym}(x,t)\rho_{sym}(x^{'},t^{'})  >_0 \mbox{      }   U^{'}_{sym}(x^{'},t^{'}) + \frac{2i}{ v_0 }\mbox{     }      ( U_{sym}(-x,t) - U^{'}_{sym}(-x,t) )  = 0
 \label{usym}
 \end{align}
 and
 \begin{align}
 \int_C dt^{'} \int dx^{'} \mbox{     }<T\rho_{asy}(x,t)\rho_{asy}(x^{'},t^{'})  >_0 \mbox{      }   U^{'}_{asy}(x^{'},t^{'})      -  \frac{2i}{ v_0 } ( U_{asy}(-x,t) - U^{'}_{asy}(-x,t))= 0
 \label{uasy}
 \end{align}
 In terms of these saddle points,
 \begin{align}
 Z[U] = e^{-\frac{1}{i v_{0}} \int_C dt \int dx \mbox{ } (U^{'}_{sym}(-x,t) - U_{sym}(-x,t)) U_{sym}(x,t) -\frac{1}{i v_{0}} \int_C dt \int dx \mbox{ } (U_{asy}(-x,t) - U^{'}_{asy}(-x,t)) U_{asy}(x,t)}
 \label{zu}
 \end{align}
 The transport through the point contact depends on the tunneling density of states (TDOS) of each edge mode. We expect the TDOS to have a power law suppression characteristic of a Luttinger liquid. The TDOS of say, the right moving edge is related to the Green's function $<T \psi^{\dagger}_{R}(x,t) \psi_{R}(x,t^{'})>$ which is evaluated using our unconventional bosonization ansatz \footnote{It could also be expressed as $<T \psi^{\dagger}_{R}(x,t) \psi_{R}(x,t^{'})> \sim < e^{ -2 \pi i \frac{1}{\nu}\int^{x} (\rho_R(y,t) + \rho_L(-y,t) )  \mbox{      }dy } \mbox{ }\mbox{ }e^{ 2 \pi i \frac{1}{\nu}\int^{x^{'}} \rho_R(y^{'},t^{'}) dy^{'}}>$, the idea is that the modified Fermi-Bose relation should be used for only one of the Fermi fields while evaluating the correlations i.e. $\lambda = 1$ for one of the operators and $\lambda=0$ for the other one (see Eq.23 in \cite{das2019nonchiral}). } as follows
 \begin{align}
 <T \psi^{\dagger}_{R}(x,t) \psi_{R}(x,t^{'})> \mbox{ }\sim\mbox{ } &< e^{ -2 \pi i \frac{1}{\nu}\int^{x} \rho_R(y,t) \mbox{      }dy }\mbox{ } \mbox{ }\mbox{ }\mbox{ }e^{ 2 \pi i \frac{1}{\nu} \int^{x^{'}} (\rho_R(y^{'},t^{'}) + \rho_L(-y^{'},t^{'}) )  \mbox{      }dy^{'} }>\nonumber \\&\sim \mbox{ }
e^{\frac{1}{2}(2 \pi i)^{2} \frac{1}{\nu^{2}}\int^{x^{'}} dy \int^{x^{'}} dy^{'} <\rho_R(y,t^{'})\rho_R(y^{'},t^{'})>} \nonumber \\
&\mbox{ }\mbox{ }\mbox{ }e^{\substack{\frac{1}{2}(2 \pi i)^{2}\frac{1}{\nu^{2}}\int^{x}dy \int^{x} dy^{'} <(\rho_R(y,t) + \rho_L(-y,t) ) (\rho_R(y^{'},t) + \rho_L(-y^{'},t) )>} }\nonumber \\&\mbox{ }\mbox{ }\mbox{ }e^{-(2 \pi i)^{2} \frac{1}{\nu^{2}} \int^{x} dy \int^{x^{'}} dy^{'} <\rho_R(y,t)(\rho_R(y^{'},t^{'}) + \rho_L(-y^{'},t^{'}) )>}
\label{rrncbt}
 \end{align}
The term that contributes to the dynamical part is
\begin{align}
e^{-(2 \pi i)^{2} \frac{1}{\nu^{2}}\int^{x} dy \int^{x^{'}} dy^{'} <\rho_R(y,t)(\rho_R(y^{'},t^{'}) + \rho_L(-y^{'},t^{'}) )>} = e^{-(2 \pi i)^{2} \frac{1}{2} \frac{1}{\nu^{2}}\int^{x} dy \int^{x^{'}} dy^{'} <\rho_{sym}(y,t) \rho_{sym}(y^{'},t^{'})> }
\end{align}
 while the other terms give only the prefactors which are constant for zero bias and are not important for extracting the TDOS exponents.
     One can write the equilibrium part of the $<\rho_{sym}\rho_{sym}>$ DDCF in momentum and Matsubara frequency space according to the transformation,
 \begin{align}
 <T\rho_{sym}(x,t)\rho_{sym}(x^{'},t^{'})>_{0}\mbox{  } = \mbox{  }
      \frac{1}{L}\sum_{q,n} e^{-iq (x-x^{'}) } \mbox{ } e^{w_n(t-t^{'}) }
      \mbox{        } <\rho_{sym}(q,n)\rho_{sym}(-q,-n)>
 \end{align}
 where $w_{n} = \frac{2 \pi n}{\beta}$ are bosonic Matsubara frequencies with $n \in \mathbb{Z}$. Writing Eq.\ref{usym} in Matsubara frequency and momentum space
 \begin{align}
 -i \beta  \mbox{  }  <\rho_{sym}(q,n)\rho_{sym}(-q,-n)>_0   \mbox{  }
  U^{'}_{sym}(q,n)  + \frac{2i}{ v_0 }   \mbox{  }
 ( U_{sym}(-q,n) -  U^{'}_{sym}(-q,n) ) = 0
 \end{align}
 where $ U_{sym}(-x,t)  \mbox{        }  =  \mbox{        }\frac{1}{-i \beta L} \sum_{q,n} e^{ i q x } e^{w_n t } \mbox{  }\mbox{  }
  U_{sym}(q,n)$.
  The generating functional for the $<\rho_{sym} \rho_{sym}>$ correlations is
  \begin{align}
  Z[U_{sym}] = e^{-  \sum_q  \mbox{  }\frac{1}{ v_{0} \beta L}
 (U^{'}_{sym}(q,-n) -  U_{sym}(q,-n))
  U_{sym}(q,n) }
  \end{align}
  The saddle point for the $U^{'}_{sym}$ field is obtained as
  \begin{align}
  U^{'}_{sym}(q,n) = \frac{2(v_{0} \beta R_{sym}(-q,n) U_{sym}(-q,n) - 2 U_{sym}(q,n))}{-4+\beta ^2 v_{0}^2 R_{sym}(-q,n) R_{sym}(q,n)}
  \end{align}
  where $R_{sym}(q,n)\mbox{  }=\mbox{   }<\rho_{sym}(q,n)\rho_{sym}(-q,-n)>_0 \mbox{ }=\mbox{ }-\frac{1}{\pi \beta v_{F}}\frac{  i v_F q}{(w_{n}-i v_{F} q)}$ is the noninteracting correlation in momentum-Matsubara frequency space. Substituting this saddle point we may write
 \begin{align}
   Z[U_{sym}] = e^{ \frac{1}{  L}\sum_{q,n}
\frac{  R_{sym}(-q,n) (2 U_{sym}(-q,n)-\beta  v_0 R_{sym}(q,n) U_{sym}(q,n))}{4-\beta ^2 v_0^2 R_{sym}(-q,n) R_{sym}(q,n)}
  U_{sym}(q,-n) }
\end{align}
But by definition the generating functional (for the most singular parts) takes the form,
\begin{align}
 Z[U_{sym}] = e^{ \frac{1}{2  L}\sum_{q,q^{'},n} <\rho_{sym}(q,n)\rho_{sym}(q^{'},-n)>   \mbox{  }   U_{sym}(q,n)
  U_{sym}(q^{'},-n) }
\end{align}
Hence we have the relation
 \begin{align}
&  \frac{1}{  L}\sum_{q,n}
\frac{  R_{sym}(-q,n) (2 U_{sym}(-q,n)-\beta  v_0 R_{sym}(q,n) U_{sym}(q,n))}{4-\beta ^2 v_0^2 R_{sym}(-q,n) R_{sym}(q,n)}
  U_{sym}(q,-n) \\& = \frac{1}{2  L}\sum_{q,q^{'},n} <\rho_{sym}(q,n)\rho_{sym}(q^{'},-n)>   \mbox{  }   U_{sym}(q,n)
  U_{sym}(q^{'},-n)
\end{align}
Now the density-density correlation function of interest in space-time domain can be calculated using,
\begin{align}
 <T\rho_{sym}(x,t)\rho_{sym}(x^{'},t^{'})> \mbox{  } = \mbox{  }
      &\frac{1}{L}\sum_{q,n} e^{-iq (x-x^{'}) } \mbox{ } e^{w_n(t-t^{'}) }
      \mbox{        } <\rho_{sym}(q,n)\rho_{sym}(-q,-n)>  \\& +       \frac{1}{L}\sum_{q,n} e^{-iq (x+x^{'}) } \mbox{ } e^{w_n(t-t^{'}) }
      \mbox{        } <\rho_{sym}(q,n)\rho_{sym}(q,-n)>
 \end{align}
 We obtain
 \begin{align}
 <T\rho_{sym}(x,t)\rho_{sym}(x^{'},t^{'})> \mbox{  } = \mbox{  }\frac{\substack{v_{0} \left(csch^{2}\left(\frac{\pi  ((t-t^{'}) v_{h}+x+x^{'})}{\beta  v_{h}}\right)+csch^{2}\left(\frac{\pi  (-(t-t^{'}) v_{h}+x+x^{'})}{\beta  v_{h}}\right)\right)\\ + \mbox{ }2 \pi  \left((v_{h}-v_{F}) csch^{2}\left(\frac{\pi  ((t-t^{'}) v_{h}+x-x^{'})}{\beta  v_{h}}\right)-(v_{F}+v_{h}) csch^{2}\left(\frac{\pi  ((t-t^{'}) v_{h}-x+x^{'})}{\beta  v_{h}}\right)\right)}}{\substack{8 \pi  \beta^{2}  v_{h}^3}}
 \label{rsymrsymint}
 \end{align}
 where the holon velocity is defined as $v_{h}^{2} = v_{F}^{2} - \frac{v_{0}^{2}}{4 \pi^{2}}$.
It is clear that in the noninteracting limit ($v_{0} \rightarrow 0$) Eq. \ref{rsymrsymint} reduces to Eq. \ref{rsymrsym}.
Now we have all the pieces in place to calculate the TDOS in presence of the point-contact.

\section{Tunneling density of states of FQHE edge states}
\label{tdos}
\subsection{Electron tunneling}
\begin{figure}
\centering
 \includegraphics[scale=0.15]{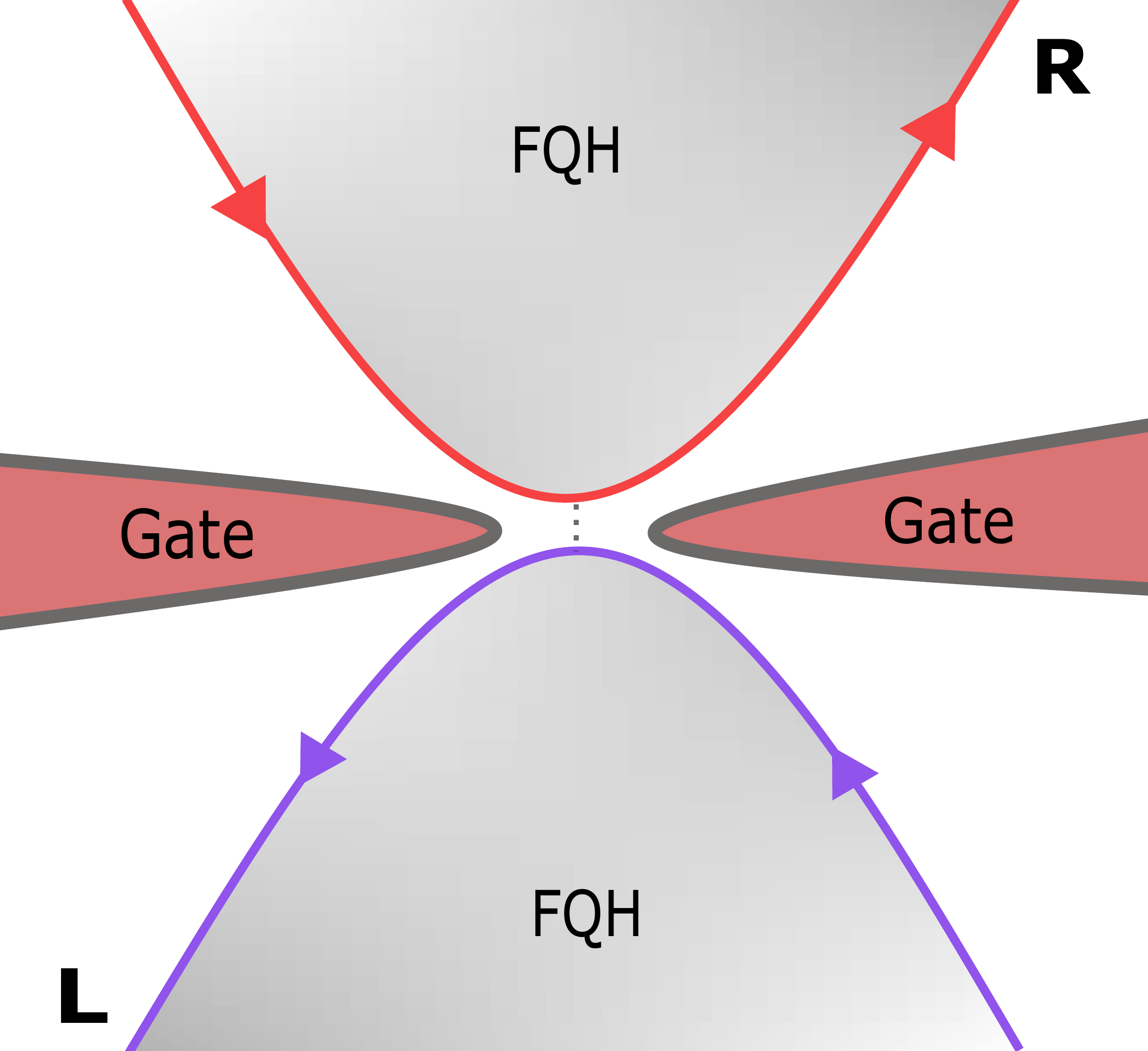}
 \caption{\small Schematic diagram of a completely pinched off point-contact geometry for electron tunneling. The point contact is controlled by electrostatic gate voltage.}
 \label{figelectun}
 \end{figure}

A gate voltage can be used to control the point contact electrostatically. Let us first consider the case where the point contact is completely pinched off such that now the tunneling between the edge states does not happen through the bulk quantum Hall fluid and the particles that are transferred between the edges during the tunneling process are actual electrons of charge $e$ (see Fig.\ref{figelectun}). Using the $<\rho_{sym} \rho_{sym}>$ correlations obtained in the previous section we calculate the TDOS at the point contact for the right-moving edge using the unconventional bosonization procedure and we obtain
\begin{align}
<T \psi^{\dagger}_{R}(0,t) \psi_{R}(0,t^{'})> \mbox{ }\sim\mbox{ } \frac{1}{sinh(\frac{\pi (t-t^{'})}{\beta})^{\frac{1}{\nu^{2}}(\frac{v_{0}+2 \pi v_{F}}{2 \pi v_{h}})}}
\end{align}
The Luttinger parameter $K$ should be equal to the filling factor $\nu$ in order to make a connection between the Luttinger liquid paradigm and FQHE edge states \cite{PhysRevLett.71.4381}. So we have
\begin{align}
K = \nu = \sqrt{\frac{\frac{v_{0}}{2 \pi  v_{F}}+1}{1-\frac{v_{0}}{2 \pi  v_{F}}}}
\end{align}
From this we obtain the relation $\frac{v_{0}}{\pi  v_{F}}=\frac{2 \left(\nu ^2-1\right)}{\nu ^2+1}$. It is easy to check that the exponent turns out to be
 \begin{align}
<T \psi^{\dagger}_{R}(0,t) \psi_{R}(0,t^{'})> \mbox{ }\sim\mbox{ } \frac{1}{sinh\left(\frac{\pi (t-t^{'})}{\beta}\right)^{\frac{1}{\nu}}}
\end{align}
After a Fourier transform the result for the TDOS at zero temperature may be written in general as
\begin{align}
D(\omega) \sim |\omega|^{m-1}
\end{align}
where $\nu = 1/m$ with $m$ an odd integer. This is a power-law with universal exponent as expected for Luttinger liquid behaviour \cite{RevModPhys.75.1449,doi:https://doi.org/10.1002/9783527617258.ch4}. For $\nu=1$ that is $m=1$ (IQHE edge state) the TDOS is a constant at the Fermi energy. For the most robust fractional quantum Hall state with $\nu=1/3$ ($m=3$) \cite{PhysRevB.23.5632} the electron tunneling density of states exponent has a value of $2$, in agreement with previous results \cite{RevModPhys.75.1449}. A similar calculation will give the same exponent for the TDOS of the left-moving edge as well.
\subsection{Quasiparticle tunneling}
\begin{figure}
\centering
 \includegraphics[scale=0.12]{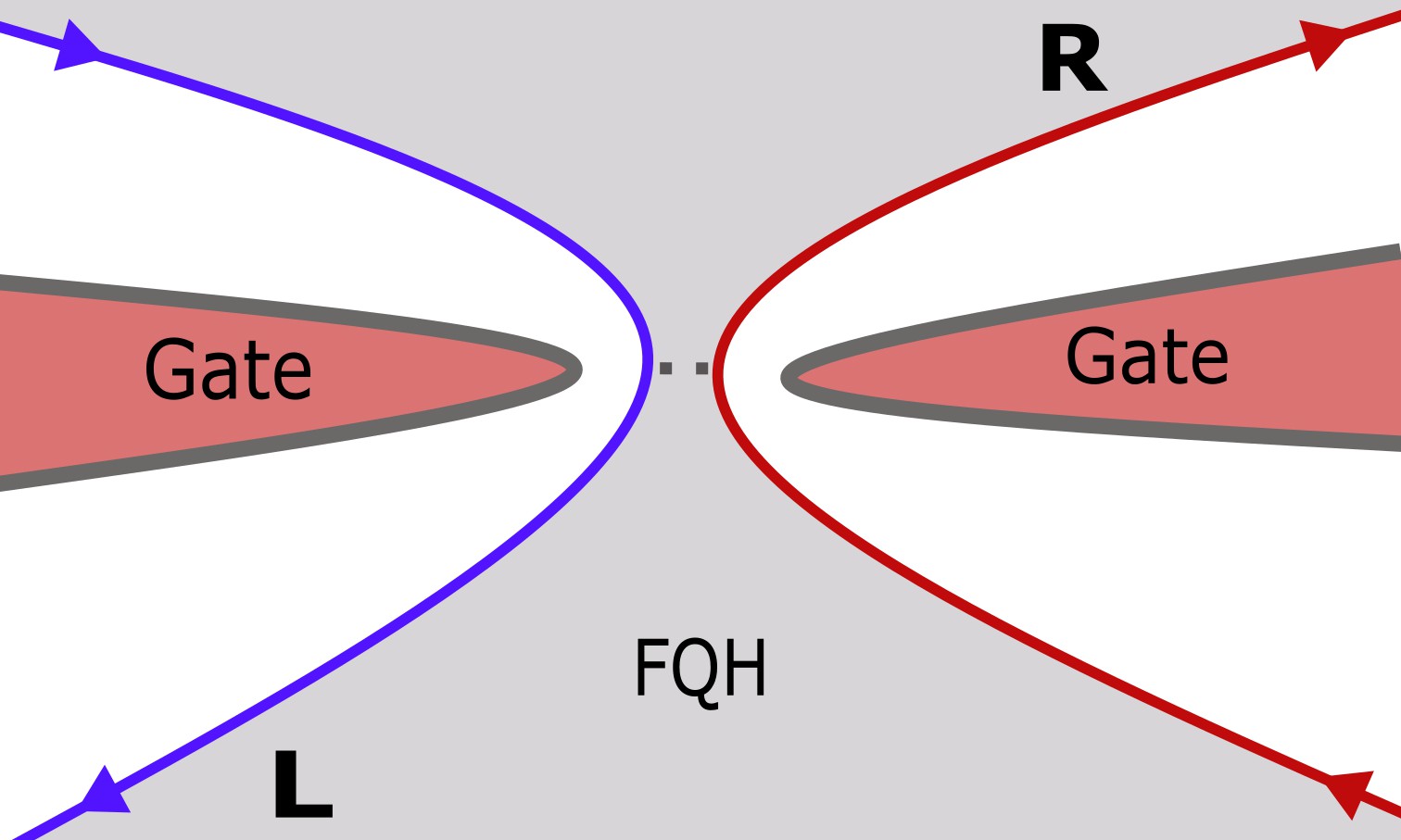}
 \caption{\small Schematic diagram of a point-contact geometry for quasiparticle tunneling through the bulk quantum Hall fluid. The point contact is controlled by electrostatic gate voltage.}
 \label{figquasitun}
 \end{figure}

Let us consider the case where the point-contact is not pinched off. In this case the particles can tunnel between the top edge and the bottom edge through the bulk quantum Hall fluid (see Fig.\ref{figquasitun}). Therefore, it is the Laughlin quasiparticles with charge $e/m$ that tunnel between edges. In this case the operators we consider are for charge $e/m$ edge excitations
\begin{align}
\psi_{\chi}(x,t) = e^{i \chi \Phi} = e^{ 2 \pi i \mbox{ }\chi\int^{x} \rho_{\chi}(y,t) dy }
\end{align}
These particles show fractional statistics, acquiring a phase factor $e^{\pm i \nu \pi}$ under exchange. The modified Fermi-Bose correspondence we consider due to the presence of the point contact is
\begin{align}
\psi_{\chi}(x,t) =  e^{ 2 \pi i \chi \int^{x} (\rho_\chi(y,t)+\lambda \mbox{ }\rho_{-\chi}(-y,t)) dy }
\end{align}
The two-point functions are evaluated in the same manner as in Eq.\ref{rrncbt} but without the $\frac{1}{\nu^{2}}$ factor appearing in the exponent. The TDOS at the point-contact for the right-moving edge is obtained as
\begin{align}
<T \psi^{\dagger}_{R}(0,t) \psi_{R}(0,t^{'})> \mbox{ }\sim\mbox{ } \frac{1}{sinh\left(\frac{\pi (t-t^{'})}{\beta}\right)^{(\frac{v_{0}+2 \pi v_{F}}{2 \pi v_{h}})}}\mbox{ }=\mbox{ }\frac{1}{sinh\left(\frac{\pi (t-t^{'})}{\beta}\right)^{\nu}}
\end{align}
In frequency space, the zero temperature TDOS takes the form
\begin{align}
D(\omega) \sim |\omega|^{\frac{1}{m}-1}
\end{align}
Unlike the case of electron tunneling the TDOS at zero energy is enhanced for the case of quasiparticle tunneling \cite{doi:https://doi.org/10.1002/9783527617258.ch4}. The TDOS for the left-moving edge can be calculated in a similar manner to obtain the same exponent.
\section{Current in response to a difference of potential between the edges}
\label{linearcurrent}
In the absence of the point contact ($\Gamma = 0$) and in the absence of interactions it follows from Eq.\ref{nobiasrelation} that  $<T\rho_{asy}(x,t)\rho_{asy}(x^{'},t^{'})>_{0}\mbox{ }=\mbox{ }<T\rho_{sym}(x,t)\rho_{sym}(x^{'},t^{'})>_{0}$. This means that in the situation without a point contact and zero bias we can compute the interacting correlations of the $\rho_{asy}$ fields following the same procedure as in Sec.\ref{ddcf}. We obtain
\begin{align}
<T\rho_{asy}(x,t)\rho_{asy}(x^{'},t^{'})> \mbox{  } = \mbox{  }\frac{\substack{-v_{0} \left(csch^{2}\left(\frac{\pi  ((t-t^{'}) v_{h}+x+x^{'})}{\beta  v_{h}}\right)+csch^{2}\left(\frac{\pi  (-(t-t^{'}) v_{h}+x+x^{'})}{\beta  v_{h}}\right)\right)\\ + \mbox{ }2 \pi  \left((v_{h}-v_{F}) csch^{2}\left(\frac{\pi  ((t-t^{'}) v_{h}+x-x^{'})}{\beta  v_{h}}\right)-(v_{F}+v_{h}) csch^{2}\left(\frac{\pi  ((t-t^{'}) v_{h}-x+x^{'})}{\beta  v_{h}}\right)\right)}}{\substack{8 \pi  \beta^{2}  v_{h}^3}}
\label{rhoasyrhoasyint}
\end{align}
 Since the right and left movers are physically separated from each other the current in response to the difference in potential between the right and left edges gives the two-terminal conductance. This current is defined as \cite{giamarchi2003quantum},
 \begin{align}
 j_{e}(x,t)\mbox{ }=\mbox{ }e v_{h} \nu (\Pi_{R}(x,t) - \Pi_{L}(x,t))\mbox{ }=\mbox{ }e v_{h} (\rho_{R}(x,t) - \rho_{L}(x,t))
 \end{align}
 where $\Pi(x,t) = \rho(x,t)/\nu$ is the conjugate momentum of the $\phi$ fields.
 We consider that a bias $V_{b}$ is applied to the right movers and the chemical potential of the left movers is taken to be zero such that $\mu_{L}-\mu_{R} = -\mu_{R} = -e V_{b} = e V$. That means we have
\begin{align}
H_{bias} = e V_{b} \int dy \rho_{R}(y,t)
\end{align}
 The average current can be written
 \begin{align}
 <j_{e}(x,t)>\mbox{ }=\mbox{ }e v_{h}\frac{<T \mbox{ }S (\rho_{R}(x,t) - \rho_{L}(x,t))>_{V_{b}=0}}{<T S>_{V_{b}=0}}
 \end{align}
 where $S = e^{-i \int_{C} dt H_{bias}}$. The current in linear response to the bias maybe computed directly
 \begin{align}
 <j_{e}(x,t)>\mbox{ }=\mbox{ }e v_{h} (-i e V_{b}) \int_{-\infty}^{\infty}dy \int_{0}^{-i \beta}dt_{1}\left(<\rho_{R}(y,t_{1}) \rho_{R}(x,t)> - <\rho_{R}(y,t_{1}) \rho_{L}(x,t)>\right)
 \end{align}
 This may be evaluated using Eqs. \ref{rsymrsymint} and \ref{rhoasyrhoasyint} and we obtain
 \begin{align}
 <j_{e}>\mbox{ }=\mbox{ }\frac{e^{2}}{h} \frac{v_{0} + 2 \pi v_{F}}{2 \pi v_{h}} V \mbox{ }=\mbox{ }\nu \frac{e^{2}}{h} V
 \end{align}
 This gives the correct expression for the fractional conductance
 \begin{align}
 G \mbox{ }=\mbox{ } \nu \frac{e^{2}}{h}
 \end{align}
 In the presence of a point contact it is not so straightforward to calculate the $<T\rho_{asy}(x,t)\rho_{asy}(x^{'},t^{'})>$ correlations with interactions. It involves a tedious procedure and if one wishes to include the bias voltage to all orders it complicates matters further. Therefore the complete results of the interacting density-density correlations and the two-point nonequilibrium Green's functions in presence of interactions, point-contact impurity and voltage bias between the edges will be discussed in a future communication.
 \section{Conclusion}
 \label{conclusion}
 We have used the modified Fermi-Bose correspondence in Eq.\ref{ncbtmod} to recover the correct tunneling density of states of FQHE edge states with a point contact. A major advantage of this approach is that the action remains quadratic in the bosonic operators even when backscattering due to the point contact impurity is included. However this method only gives the most singular parts of the correlation functions correctly. This quadratic (bosonic) action is used to derive the correlation functions in presence of mutual interactions using the generating functional method in Sec.\ref{ddcf}. In conventional bosonization, the backscattering between the right and left movers due to the impurity produces a boundary sine-Gordon model that is not quadratic in the bosonic fields \cite{saleur1998lectures}. But in this alternative approach, the modified Fermi-Bose correspondence encodes the backscattering while leaving the action quadratic even with impurity and mutual (forward scattering) interaction between the fermions present. This method gives the most singular parts of the correlation functions of inhomogeneous Luttinger liquids exactly \cite{Danny_Babu_2020}. Our goal is to obtain the most singular parts of the non-equilibrium Green's function for this system exactly with a non-perturbative treatment of the point contact impurity. This is crucial in order to obtain the non-Ohmic tunneling I-V characteristics and express the conductance as a universal scaling function for a generic impurity strength. It involves calculating the $<\rho_{asy}\rho_{asy}>$ correlations out of equilibrium with interactions and in presence of a point contact. This will be considered in a subsequent paper.\\
 To summarize, this unconventional approach to bosonization enables the calculation of the most singular parts of the correlation functions of fermions with localized scattering centers where backscattering takes place and in presence of mutual fermion-fermion forward scattering interactions to be expressed in terms of elementary functions of position and time. This is accomplished by verifying, a-posteriori, the following claims:
 \\ \mbox{  } \\
(i)\textit{ Fact I:} The conventional bosonization scheme of the above mentioned model leads to a Lagrangian density that is nonlinear (Sine-Gordon) \cite{saleur1998lectures} but has a spatially local expression. Because of the Sine-Gordon term, it is not soluble.
 The most singular parts of the mutual (auto) correlations of density fluctuations involves realising (with proof \cite{Danny_Babu_2020}) that all the odd moments of the density fluctuations of the above system vanish identically whereas all the higher order even moments although non-zero are less singular than the leading second order moment and therefore ignorable.
\\ \mbox{  } \\
(ii) \textit{Claim I:} The most singular part of the above Lagrangian density in \textit{Fact I} that recovers the density density correlation correctly and
yields the result that all higher moments vanish (they are less singular so we feel entitled to ignore them) is  spatially non-local but purely quadratic in the bosons and therefore soluble.
 \\ \mbox{  } \\
 (iii) \textit{Fact II:} The conventional bosonization scheme has a Fermi-Bose correspondence that is an operator identity that is tied to the translationally invariant fermion basis.
\\ \mbox{  } \\
(iv) \textit{Claim II:} The main message is that it is possible to study backward scattering from a stationary impurity in such a way that \textbf{(only) the most singular parts} of the correlation functions can be written down exactly using a modified bosonization framework that transfers part of the burden of shouldering the Sine Gordon character of the conventional approach to the modified Fermi Bose correspondence while retaining the action of the theory quadratic (albeit spatially non-local) in the bosons and therefore, soluble (see \hyperref[AppendixA]{Appendix A}).
 \\ \mbox{ } \\ The main contribution of this work is to show that important physical attributes such as tunneling density of states exhibit universal power law behavior consistent with other well-established approaches found in the literature, even as the (most singular parts of the) full  space-time Green functions - which only the present approach can calculate - remain non-universal. 

 \section*{Author contributions}
All authors contributed equally to the paper. All the authors have read and approved the final manuscript.
Both authors contributed equally to the analytical calculations contained in the present manuscript.\\
\mbox{ }\\

\section*{Declarations}
\textbf{Conflict of interest}\mbox{ }The authors have no competing interests.
%--------------------------------------------------------------------------------------------------
%----------------------------------------APPENDIX

\section*{APPENDIX A: Most singular part of the action}
\label{AppendixA}
\setcounter{equation}{0}
\renewcommand{\theequation}{A.\arabic{equation}}
The densities $ \rho_{sym},\rho_{asy} $ are defined such that they are decoupled
  $ <\rho_{sym} (x,t)\rho_{asy}(x^{'},t^{'}) > = 0 $ even when there is backscattering at the impurity. The full action can be written as,
\[
S = S_{sym}[\rho_{sym}] + S_{asy}[\rho_{asy}]
\]
The crucial claim (proved in \cite{Danny_Babu_2020}) is that the generating function of the current and density include only the most singular quadratic terms. We define,
\[
Z_0[U_{sym}, U_{asy}] =  \int \mbox{     }  D[\rho_{sym}] D[\rho_{asy}] \mbox{   } \mbox{  }
e^{ i S[\rho_{sym},\rho_{asy}] } \mbox{  }
e^{ \int dx \int_c dt \mbox{  } \rho_{asy}(x,t) U_{asy}(x,t) + \int dx \int_c dt \mbox{  } \rho_{sym}(x,t) U_{sym}(x,t)   }
\]
Given that we are only including the most singular parts of the correlation functions (all higher odd moments anyway vanish identically, but the even ones are less singular than the leading quadratic terms included), the form of $ Z_0 $ is the exponential
 of a quadratic form in $   \rho_{sym}, \rho_{asy} $.
\[
Z_{0,most-sing}[U_{sym}, U_{asy}]  \mbox{           } = \mbox{              }
e^{ \frac{1}{2} \int dx \int dx^{'} \int dt \int dt^{'} \mbox{  }\sum_{a = asy,sym } \mbox{   }
<T\mbox{  }\rho_a(x,t)\rho_a(x^{'},t^{'}) >_0 \mbox{ } U_{a}(x,t) U_{a}(x^{'},t^{'}) } \mbox{ } \mbox{  }
\]
These quantities viz. $   <T\mbox{  }\rho_a(x,t)\rho_a(x^{'},t^{'}) >_0 $ maybe derived explicitly using elementary Fermi algebra and are written down below:
\begin{align}
 <\rho_{sym}(x,t)\rho_{sym}(x^{'},t^{'})>_{0} \mbox{ }=\mbox{ }2\left[    \frac{i}{2\pi} \frac{ \frac{ \pi }{\beta v_F } }{\sinh( \frac{ \pi }{\beta v_F } (x-x^{'}-v_F(t-t^{'}) ) )  } \right]^2
 \label{rsymnonint}
 \end{align}
 \begin{align}
 <\rho_{sym}(x,t)\rho_{asy}(x^{'},t^{'})>_{0} \mbox{ }=\mbox{ } 0
 \end{align}
 \begin{align}
 <\rho_{asy}(x,t)\rho_{asy}(x^{'},t^{'})>_{0} \mbox{ }=\mbox{ }&\frac{-((4 v_{F}^{2} + \Gamma^{2})^{2}-32 v_{F}^{2} \Gamma^{2} \theta(x))((4 v_{F}^{2} + \Gamma^{2})^{2}-32 v_{F}^{2} \Gamma^{2} \theta(x^{'}))}{2 v_{F}^{2} \beta^{2}(4 v_{F}^{2} + \Gamma^{2})^{4}\sinh( \frac{ \pi }{\beta v_F } (x-x^{'}-v_F(t-t^{'}) ) )^{2}}\nonumber \\ &-\frac{64(-4 v_{F}^{3} \Gamma + v_{F} \Gamma^{3})^{2} \cos(\frac{- e V_{b}(x-x^{'}-v_F(t-t^{'}) )}{v_{F}})\mbox{ }\theta(x)\theta(x^{'})}{2 v_{F}^{2} \beta^{2}(4 v_{F}^{2} + \Gamma^{2})^{4}\sinh( \frac{ \pi }{\beta v_F } (x-x^{'}-v_F(t-t^{'}) ) )^{2} }
 \label{rhoasy}
 \end{align}
This means the most singular part of the action is spatially non-local but quadratic in the boson terms which we may formally write,
\[
e^{ i S_{most-sing}[\rho_{sym},\rho_{asy}] } \mbox{   }  =   \mbox{     } \int D[U_{sym}] \int D[U_{asy}] \mbox{  }\mbox{  }
 \mbox{              }
e^{ \frac{1}{2} \int dx \int dx^{'} \int dt \int dt^{'} \mbox{  }\sum_{a = asy,sym } \mbox{   }
<T\mbox{  }\rho_a(x,t)\rho_a(x^{'},t^{'}) >_0 \mbox{ } U_{a}(x,t) U_{a}(x^{'},t^{'}) } \mbox{ } \mbox{  }
\]
\[
\times \mbox{   }e^{ -\int dx \int_c dt \mbox{  } \rho_{asy}(x,t) U_{asy}(x,t) - \int dx \int_c dt \mbox{  } \rho_{sym}(x,t) U_{sym}(x,t)   }
\]
It is this action we are (indirectly) using in lieu of the full Sine-Gordon action. While the full action correctly captures all aspects of the model, it is not soluble. The effective quadratic (albeit spatially non-local) action above only gives the most singular parts of the correlation functions of the theory provided in addition we also modify the Fermi-Bose correspondence to shoulder part of the burden of capturing the effect of back-scattering at the impurity in the manner we have described in the manuscript.
Thus the claim is 
\[
[ S_{full = Sine-Gordon} ]_{most-singular-part} \mbox{  } = \mbox{  } S_{spatially-nonlocal-but-quadratic-and-soluble}
\] 
and
\[
[ \psi_R(x,t)  ]_{most-singular-part} \mbox{  } = \mbox{  } [ e^{ 2\pi i \int^{x} dy \mbox{  } \rho_R(y,t) } ]_{most-singular-part}
 \mbox{  } = \mbox{  } [ e^{ 2\pi i \int^{x} dy \mbox{  } ( \rho_R(y,t) + \lambda \mbox{  } \rho_L(-y,t) ) } ]_{\lambda = 0,1}
\]
The modified Fermi-Bose correspondence rather than being tied to the inappropriate spatially homogeneous basis now encodes the effect of back-scattering at the impurity through the additional term in the exponent (while calculating the two point function $ \lambda = 1 $ should be used at most once, the reason is explained in an earlier published work (point splitting considerations) \cite{das2019nonchiral}.  

%\newpage
\section*{References}
\bibliographystyle{iopart-num}
\bibliography{refnew}
\end{document}